\documentclass[preprint,11pt]{elsarticle}

\usepackage{amssymb}
\usepackage{amsmath}

\usepackage{dsfont}
\usepackage{tikz}
\usetikzlibrary{bayesnet}
\usetikzlibrary{positioning,arrows.meta,shapes,calc}
\usepackage{makecell}
\usepackage{booktabs}
\usepackage{pdflscape}
\usepackage{hyperref}
\usepackage{float}
\usetikzlibrary{arrows}

\usepackage[margin=1.3in, includeheadfoot]{geometry}
\usepackage[utf8]{inputenc}
\usepackage{centernot}
\newcommand{\indep}{\perp \!\!\! \perp}
\newcommand{\notindep}{\centernot{\indep}}

\journal{Knowledge-Based Systems}

\begin{document}

\begin{frontmatter}



\title{Structural Refinement of Bayesian Networks for Efficient Model Parameterisation}


\author[affil1]{Kieran Drury\corref{cor1}} 
\ead{kieran.drury@warwick.ac.uk}
\author[affil1]{Martine J. Barons} 
\author[affil1]{Jim Q. Smith} 

\cortext[cor1]{Corresponding author}

\affiliation[affil1]{organization={Department of Statistics, University of Warwick}, 
            city={Coventry},
            postcode={CV4 7AL},
            country={UK}}

\begin{abstract}
Many Bayesian network modelling applications suffer from the issue of data scarcity. Hence the use of expert judgement often becomes necessary to determine the parameters of the conditional probability tables (CPTs) throughout the network. There are usually a prohibitively large number of these parameters to determine, even when complementing any available data with expert judgements. To address this challenge, a number of CPT approximation methods have been developed that reduce the quantity and complexity of parameters needing to be determined to fully parameterise a Bayesian network. This paper provides a review of a variety of structural refinement methods that can be used in practice to efficiently approximate a CPT within a Bayesian network. We not only introduce and discuss the intrinsic properties and requirements of each method, but we evaluate each method through a worked example on a Bayesian network model of cardiovascular risk assessment. We conclude with practical guidance to help Bayesian network practitioners choose an alternative approach when direct parameterisation of a CPT is infeasible.

\end{abstract}




\end{frontmatter}



\section{Introduction}
Bayesian networks (BNs) are a long-established probabilistic graphical modelling tool for intuitively capturing complex, real-world systems \citep[see e.g.][]{Pearl1988ProbReasoning,JensenNielsen2007,KorbNicholson2011}. They are used in a variety of domains such as environmental risk assessment \citep{Kaikkonen2020}, clinical decision support in healthcare \citep{Kyrimi2021ScopingReview}, neuroscience \citep{BielzaLarranaga2014}, cyber security \citep{Chockalingam2017} and terrorism intervention \citep[see e.g.][]{Mohsendokht2024,DrurySmith2024}. Within these domains, BNs are used as descriptive, predictive and prescriptive models, demonstrating the wide applicability and breadth of BNs as a modelling technique. BNs further benefit from the intuitive graphical structure they possess, and the natural simplicity with which their outputs can be expressed. Particularly in an era of complex, black-box machine learning and artificial intelligence methodologies, model transparency, interpretability and explainability is a desired and often demanded feature of any model whose outputs are to be used in the real world \citep{Jobin2019,Balasubramaniam2023,Goodman2017}. Bayesian networks are intuitive and explainable by design, yet can model highly complex systems, and therefore provide an ideal solution to the performance-explainability trade-off which is so often an issue in the modern AI world \citep{Derks2024}. This benefit of BNs is amplified further when BN models are constructed to meet published guidelines for transparency and reproducibility \cite[see e.g.][]{Barons2025}.

One key drawback, however, of Bayesian network modelling is the number of parameters that need to be determined to fully parameterise a model \citep{Rohmer2020}. One typical method for parameterising a BN using data is through computing relative frequencies of each variable's possible states given its set of predictor variables. The challenge is that the vast number of parameters to be determined this way requires datasets with an incredibly large number of observations to ensure each parameter estimate is reliable. This amount of data is often not available to the modeller \cite{Rohmer2020}. This data scarcity issue is further accompanied by other data quality issues of sparsity, missingness, irrelevance, obsolescence and sampling biases, among others \cite{Priestley2023}. Therefore, even when some data \textit{is} available, it may be riddled with inadequacies that significantly impede the reliability of the BN modelling outputs. In many cases, relevant, high-quality data may not even exist or be accessible to start with.

For many applications, it is thus insufficient to rely solely on existing data for parameterising a BN. One immediate option would be to set out to collect the required data, ensuring its sufficient quality and quantity. For any moderate-to-large BN, this would be extremely resource intensive. Data collection is not likely to be a feasible option in many cases. The more viable alternative in this scenario is to call on expert judgement to aid the construction of the model. Not only is expert judgement able to be used to parameterise the network, but it is also commonly used to determine the structure of the network \citep{Kitson2023} - a task that has even heavier data requirements than parameter learning. In this paper, we focus on the use of expert judgement for Bayesian network parameterisation in the context of scarce or unavailable data.

Several elicitation methodologies have been developed that structure and support the elicitation of probabilistic judgements from groups of experts, including the IDEA protocol \citep{Hanea2017IDEA} and the Sheffield Elicitation Framework (SHELF) \citep{Gosling2018}. While these methods help mitigate the effects of cognitive biases and other issues surrounding elicitation that we discuss in Section \ref{BNs}, the following two problems persist. The first is the \textit{quantity} of parameters that need eliciting, and the second is their \textit{complexity}. These issues can be circumvented through reducing the dimensions of the parameter space of the BN. There are several ways in which this can be done. Some such methods focus purely on quantitative rules such as regression and interpolation. Other methods focus on refining the structure of the network to achieve this goal. This paper explores and reviews a variety of structural approaches that facilitate efficient yet faithful parameterisation of a Bayesian network.

The paper is laid out as follows. Section \ref{BNs} provides a more detailed introduction to Bayesian networks and their elicitation through expert judgement. Section \ref{Example} introduces our running example of a Bayesian network that models cardiovascular disease risk factors. This cardiovascular BN model has been parameterised through a suitably large dataset, thus providing a suitable benchmark model with which to test the structural methods discussed in this paper. Section \ref{Methods} introduces each of these structural methods, demonstrating their implementation and discussing their characteristics. Section \ref{Comparison} features a practical comparison of these methods through a worked example, including some suggestions on when each method may be suitable to use. The paper concludes with a brief discussion about the use of these structural methods in practical BN modelling problems.

\section{Bayesian Network Parameterisation Under Scarce Data} \label{BNs}

\subsection{Bayesian Network Structure}
A Bayesian network is a probabilistic graphical model representing a system of variables through a set of interconnected nodes $\mathbf{X}$. Two nodes are connected by a directed edge whenever there may be a probabilistic dependence between the two nodes. Edges are determined by encoding a set of \textit{conditional independence statements} of the form $\mathbf{X}_A\indep \mathbf{X}_C\mid\mathbf{X}_B$, where each component is a subset of $\mathbf{X}$. When such a conditional independence statement holds, it must be the case that every path from a node in $\mathbf{X}_A$ to a node in $\mathbf{X}_C$ is \textit{blocked} or \textit{d-separated} by the nodes forming $X_B$ \citep{Pearl2009,lauritzen1996graphical}. The set of conditional independence statements to be encoded often stems from \textit{irrelevance statements} that domain experts provide \cite{Smithbook2010}. Edges are drawn into the network resulting from these irrelevance statements such that the set of conditional independence statements implied is faithful to the elicited irrelevance statements. We often draw these arrows to represent causal rather than correlational information flow, especially when the BN is to be used to model interventions \cite{Pearl2009}. These arrows must be drawn to ensure the structure of the network forms a \textit{directed acyclic graph} (DAG) \cite{Pearl2009}. 

\begin{figure}[h!]
    \centering
    \scalebox{1.25}{\begin{tikzpicture}
     \node[latent] (X3) {$X_3$};%
     \node[above=of X3] (empty) {}; %
     \node[latent,left=1cm of empty] (X1) {$X_1$}; %
     \node[latent,right=1cm of empty] (X2) {$X_2$}; %
     \node[below=of X3] (empty2) {};
     \node[latent,left=1cm of empty2](X4){$X_4$};
     \node[latent,right=1cm of empty2] (X5) {$X_5$}; %
     \edge {X1,X2} {X3};
     \edge {X3}{X4,X5};
    \end{tikzpicture}}
    \caption{Example DAG structure of a Bayesian network on five nodes}
    \label{fig:ExampleDAG}
\end{figure}
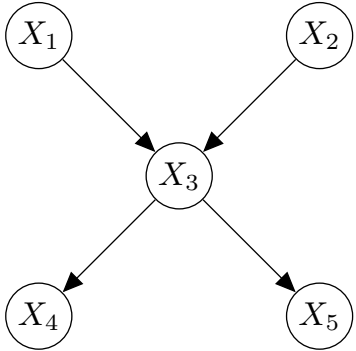

An example Bayesian network structure on five nodes is shown in Figure \ref{fig:ExampleDAG}. $X_1$ and $X_2$ are \textit{root nodes} that both have one child, $X_3$, which is itself a \textit{parent} of the \textit{leaf nodes} $X_4$ and $X_5$. $X_1$ and $X_2$ are ancestors of $X_3$ (as parents) and of $X_4$ and $X_5$ (as grandparents). Similarly, $X_4$ and $X_5$ are descendants of $X_1$, $X_2$ and $X_3$. It is simple to verify that the network structure is a DAG as no cycles are present. In Figure \ref{fig:ExampleDAG}, we have $X_4\indep X_5\mid X_3$ because all paths (ignoring directionality) between $X_4$ and $X_5$ \textit{are} blocked by $X_3$, but $X_1\notindep X_2 \mid X_3$ because the path from $X_1$ to $X_2$ through $X_3$ is a collider and is thus opened by $X_3$ (see e.g. \cite{Pearl2009,lauritzen1996graphical}).

\subsection{Conditional Probability Tables}
Having determined the structure of a Bayesian network, the next stage is to quantify the model by parameterising each of the dependencies throughout the network. For a discrete BN - in which every node has a finite number of states - this is performed by specifying a conditional probability table (CPT) for each node in the network. Where a node has no parents, this amounts to simply specifying a marginal distribution over the node's states. Where a node \textit{does} have parents, a conditional probability distribution (CPD) over the child's states is specified for each configuration of parent node states. Each row of the CPT corresponds to a unique configuration of parent states alongside a CPD for the child node. Let $s_i$ denote the number of states of each of the $n$ parents of the child node $Y$, and $s_c$ that of the $Y$ itself. The number of parameters to be determined to fully define the CPT of $Y$ is given by:
\begin{equation}\label{CPTParameterEq}
N_{Y}=\left(\prod\limits_{i=1}^{n} s_i\right)\cdot\left(s_c-1\right)
\end{equation}

A general example of a CPT is shown in Table \ref{fig:ExampleCPT}. If we constrain $X_1$, $X_2$ and $X_3$ from Figure \ref{fig:ExampleDAG} to be binary, then the CPT in Table \ref{fig:ExampleCPT} reflects the parameterisation of the local dependencies influencing $X_3$. The number of states a node has, together with the number of states of each of its parents, can be denoted concisely by the \textit{local state structure}, written as $(s_1,s_2,\ldots,s_n)\rightarrow s_c$. The local state structure for this example is given by $(2,2)\rightarrow 2$. We can see in this example CPT that we have four ($s_1\cdot s_2$) combinations of parent values, with each row's CPD parameterised by just one free parameter in the interval $[0,1]$. This CPT therefore requires four parameters for it to be fully defined, as expected following Equation \ref{CPTParameterEq}. In general, we can denote these parameters by $p_{c(k)}$ where $c$ refers to the child state and $k$ the row of the CPT.

\begin{table}[h]
    \centering
    \caption{General conditional probability table for the local state structure $(2,2)\rightarrow2$}
    \label{fig:ExampleCPT}
    \vspace{0.5em}
    \begin{tabular}{!{\vrule width 1.4pt}c|c|c|c!{\vrule width 1.4pt}}
        \noalign{\hrule height 1.4pt}
        $X_1$ & $X_2$ 
        & $\mathbb{P}(X_3=0\mid \text{pa}(X_3))$ 
        & $\mathbb{P}(X_3=1\mid \text{pa}(X_3))$  \\
        \hline
        $0$ & $0$ & $p_{0(1)}$      & $1-p_{0(1)}$ \\
        $0$ & $1$ & $p_{0(2)}$    & $1 - p_{0(2)}$ \\
        $1$ & $0$ & $p_{0(3)}$    & $1 - p_{0(3)}$ \\
        $1$ & $1$ & $p_{0(4)}$      & $1-p_{0(4)}$ \\
        \noalign{\hrule height 1.4pt}
    \end{tabular}
\end{table}

\subsection{The Need for Elicitation}
While the CPT in Table \ref{fig:ExampleCPT} only requires four parameters to be determined, many BNs developed for real-world applications feature CPTs containing a far greater amount of parameters. As Equation \ref{CPTParameterEq} demonstrates, the number of entries in a given CPT grows exponentially with the number of parents the child node has, and polynomially (with degree determined by the number of parents whose number of states we vary) in the number of states each parent has. Across even a relatively small BN, the total quantity of parameters to be determined can pose a significant challenge for producing reliable estimates.

Bayesian network parameterisation is typically performed using data-driven algorithms where possible. A simple data-driven method is to take relative frequencies of each child outcome $y$ conditional on each configuration of parent values $\mathbf{x}$. In this way, the CPT can initially be constructed as a contingency table before normalising the rows to obtain probabilities. Where $n_{c(k)}$ denotes the frequency of $Y=c$ for row $k$ in the contingency table, and $n_{(k)}$ the total number of observations for that row, the CPT parameters are given by:
\begin{equation}
\hat{p}_{c(k)}=\frac{n_{c(k)}}{n_{(k)}}    
\end{equation}
This corresponds to finding the maximum likelihood estimator (MLE) of each CPT parameter \citep{Rohmer2020}. However, when accounting for the large number of parameters we are estimating across the network, this approach requires a very large dataset to ensure an acceptable degree of stability in the parameter estimates. It is likely that, in many modelling applications, the available data will not provide a sufficiently high number of observations for every configuration of the variables in each local structure, leading to unstable and unreliable estimates \citep{Rohmer2020}. Furthermore, the dataset used to parameterise a node must jointly record all its parent variables, and the definitions of these variables must align with the current modelling objectives. The data must also meet general quality criteria such as relevance, timeliness and cleanliness (see e.g. \cite{Priestley2023}). In many domains such as volcanology \citep{Christophersen2018}, maritime accident prevention \citep{Hanninen2014}, human reliability analysis \citep{Podofillini2023}, cyber security \citep{Chockalingam2017} and ecosystem services modelling \citep{Landuyt2013}, such data is often simply not available.

In some limited cases, it may be feasible to collect reliable, primary data to support the learning of Bayesian network parameters. This typically depends on the application domain and the intended scope of the model. Some cases of Bayesian network modelling through the collection of primary data can be seen in the healthcare domain \citep{vanderStap2022,Yin2015} where collection of patient data through surveys is routine. Such cases are far rarer in other domains. However, even within the healthcare domain, there are still many concerns about the robustness of the data collection process \cite{Constantinou2016}. Even with the ability to collect primary data, data inadequacy is given as a major barrier to the increased adoption of BNs for medical research \citep{Kyrimi2021}, hence expert judgement is still often integrated into medical BN models \cite{Polotskaya2024}.

It is often not possible to rely exclusively on existing data, or the collection of primary data, for BN parameterisation, or for the even more data-hungry task of learning the network structure. The primary solution is to call upon expert judgement to support the construction of the model. Below we focus on the elicitation of the BN parameters rather than its structure; in this paper, we assume that the network structure is known. Guidance for the elicitation of the structure of a BN \cite{Burgman2021} and about learning the BN structure through data \cite{Kitson2023} is beyond the scope of this paper.

\subsection{Quantitative Elicitation Approaches}
When data is available but not in sufficient quantity to ensure stable parameter estimates, it is possible to utilise expert judgement to complement this data. This is often done through the elicitation of a Dirichlet prior that can then be updated through any data that is available, or through new data that becomes available. The elicited Dirichlet prior is conjugate to the multinomial data that is often used for standard Bayesian prior-to-posterior updating, ensuring that we arrive at a posterior distribution that is also Dirichlet \citep[see e.g.][]{Smithbook2010}. Each CPT parameter can then be estimated through maximising the likelihood of this posterior distribution in a process called \textit{maximum a posteriori} (MAP) estimation \citep[see e.g.][]{Zhou2014,Rohmer2020}. A number of additional methods for integrating data and expert judgement for BN parameterisation, in particular those based on expert-elicited qualitative parameter constraints, can be found in other literature on the topic \cite[see e.g.][]{Zhou2014,Rohmer2020}. 

Sometimes there is such little high-quality data available that expert judgement becomes the sole source of information for BN parameterisation. In a review of published Bayesian network models for environmental risk assessment \citep{Kaikkonen2020}, 18 out of the 69 models (for which the source of the parameter estimates was specified) utilised expert judgement without any integrated data-driven parameter learning techniques. This compares to 41 out of 69 models that utilised expert judgement in combination with some level of data-driven learning. Similarly, a review of BN models for ecosystem service modelling \citep{Landuyt2013} revealed 13 out of 44 models (that specified use of data or expert judgement) used expert judgement without any data learning. 23 of the 44 models used a combination of expert judgement and data, while only 8 models exclusively used data-driven approaches. To ensure accuracy and consistency of the responses elicited from domain experts, and to ensure that any model constructed through expert elicitation is constructed transparently, it is important to develop and utilise elicitation methodology that is carefully structured.

Several structured expert judgement (SEJ) methodologies have been developed and widely utilised since the mid-twentieth century. The earliest of these is the Delphi method which revolves around anonymity between experts, iterative rounds of controlled group feedback and mathematical aggregation of final responses where consensus is not naturally met \citep{DalkeyHelmer1963,Cooke1991,ohaganbook}. Many modifications have been made to the original Delphi method since its inception \cite{Cooke1991,ohaganbook}, and practical considerations for the use of these Delphi methods are long established \cite{RoweWright2001}. A more recent SEJ methodology is the Sheffield Elicitation Framework (SHELF) \cite{Gosling2018}. SHELF is built upon group discussions guided by a facilitator who aims to encourage the group towards a consensus. Experts only provide their estimates \textit{after} group discussions have taken place. It incorporates aspects of mathematical aggregation, the output of which is shared and discussed with experts, allowing them to make modifications until all experts are satisfied. Another widely used SEJ methodology is the IDEA protocol \citep{Hanea2017IDEA}, standing for \textit{Investigate}, \textit{Discuss}, \textit{Estimate} and \textit{Aggregate}. It encourages experts to individually investigate a quantity of interest before providing a private first-round estimate. The experts then meet for a group discussion, enabling the sharing of evidence, opinions and reasoning. After this, experts may privately revise their initial estimates to form their final responses which are mathematically aggregated to obtain an overall estimate for the quantity of interest. Further practical considerations and details of this method's implementations can be consulted elsewhere \cite{Hemming2018IDEAGuide}.

These above methods, when applied carefully and thoroughly, are generally accepted to facilitate a faithful elicitation of experts' probabilistic assessments which can then be integrated into the Bayesian network modelling paradigm. Through the use of these methods, the complexity of the required judgements somewhat decreases as experts are taken through the process with a high degree of guidance. However, the inherent complexity associated with assessing probabilities, especially those featuring multiple conditioning variables, still remains. Furthermore, these methods do not reduce the vast number of probabilities needing to be elicited across a network, and SEJ methodologies can be highly resource-intensive even when eliciting just a relatively small number of probabilities. This issue is even formally acknowledged by the UK Government in guidance on high-quality analysis in which it is stated that ``formal expert elicitation is costly in time and resource" \citep{AquaBook}. It proceeds to explain that less formal methods should be utilised to provide initial estimates through which target variables should be selected for more formal elicitation.

Our research focuses on this point - the development and use of less formal elicitation methodology that nonetheless remains faithful to expert beliefs and still guards against cognitive biases. These ``less formal methods" in the context of Bayesian network parameterisation include methods that approximate CPTs using fewer, less complex judgements than formal elicitation requires. We classify these approaches, which typically reduce both the quantity and the complexity of judgements simultaneously, into structural methods - the focus of this paper - and purely quantitative methods such as regression and interpolation.

Before we proceed to focus on these structural approaches, we first highlight important work providing alternative, non-structural approaches to the problem of efficient BN parameterisation. An analysis of three particular quantitative methods - namely InterBeta \citep{Mascaro2022}, the Ranked Nodes Method \citep{Fenton2007} and the Functional Interpolation Method \citep{Podofillini2015} - can be found in \cite{Blomaard2025}. Further, we highlight a review \cite{Mkrtchyan2016} that evaluates the Functional Interpolation Method \citep{Podofillini2015}, the Ranked Nodes Method \cite{Fenton2007}, the Cain Calculator \citep{Cainsmethod}, Wisse's EBBN Method \citep{Wisse2008} and R\o{}ed's Hybrid Causal Logic Method \citep{Roed2009}. These methods employ a mixture of approaches, including direct interpolation between anchor CPT rows, parametric interpolation of distributions fitted to anchor rows, and weighted aggregation of parent node values. A number of additional quantitative methods lie outside the scope of the above review, including Hassall's algorithm \citep{Hassall2019}, Phillipson's methods \citep{Phillipson2021}, Das' Weighted Sum Algorithm \citep{Das2004} and Kemp-Benedict's Influence Weights and Likelihood Methods \citep{Kemp-Benedict2008}.

The above quantitative approximation methods can, and often should, be used in conjunction with the below structural methods. Indeed, these structural methods simply aim to reduce the parameter space of a given CPT, and thus do not specify a complete quantitative approximation of any CPT. While the reduction of the parameter space may enable more efficient formal, direct elicitation of the parameters within the approximate CPT, this elicitation may be tricky as elements of the real-world system may be forgotten during the refinement of the local structure of a given node. Therefore, it may be appropriate or even necessary to utilise a quantitative approximation method in combination with the particular choice of structural approach.

\section{Cardiovascular Bayesian Network Example} \label{Example}
The Cardiovascular Bayesian network \citep{Ordovas2023} is a recently developed model of cardiovascular diseases (CVD), available through the `bnRep' R package \citep{bnRepLeonelli2025}. CVD accounts for over 45\% of all deaths across Europe, providing the motivation to develop a state-of-the-art predictive model that can be used as a decision-support tool to support diagnosis and treatment \citep{Ordovas2023}. The model referenced includes a variety of CVD risk factors (CVRFs) as established by the World Health Organisation (WHO), categorised as modifiable CVRFs and non-modifiable CVRFs, as well as other linked medical conditions. 

The model structure and its CPTs were learnt through a large dataset of almost one million records extracted from annual health assessments of working adults with private health insurance in Spain. This dataset was then combined with census information to integrate data on socioeconomic status and education level. After removal of outliers, duplicates and rows with missing values or recording errors, the dataset contained 205,087 records from between 2012 and 2016 \cite{Ordovas2023}.

After discretisation of continuous variables, the discrete BN structure was developed - initially through the greedy thick thinning algorithm, and later refined by three CVD experts \cite{Ordovas2023}. The CPT parameters were learnt through the use of a standard multinomial-Dirichlet model with uniform priors. The authors report that this did not lead to any clear data availability issues due to the large dataset used. A model validation process was also performed \cite{Ordovas2023}. The BN structure is shown in Figure \ref{fig:CardiovascularBN}:

\begin{figure}[h!]
    \centering
    \includegraphics[width=0.6\linewidth]{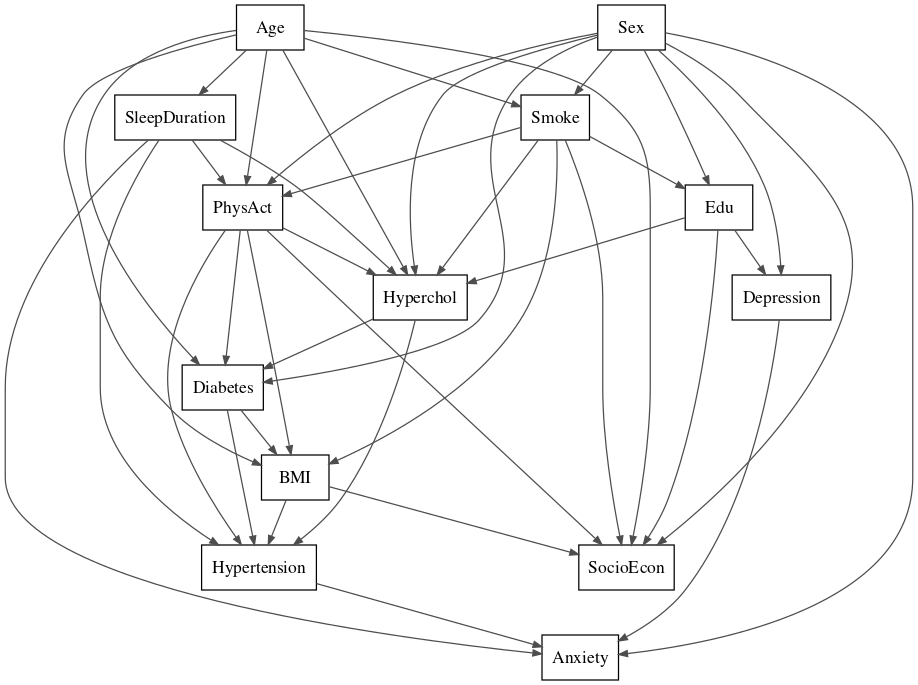}
    \caption{Cardiovascular Bayesian network \cite{Ordovas2023}}
    \label{fig:CardiovascularBN}
\end{figure}

We use this Bayesian network model as a worked example to compare the structural methodologies discussed in this paper. We do so because it is a recently developed, published BN that is readily available through the `bnRep' R package \cite{bnRepLeonelli2025}, and as its CPTs have been parameterised through a sufficiently large dataset. Crucially, it also features multiple nodes that have at least four parents, providing a suitably complex environment in which to test each structural methodology.

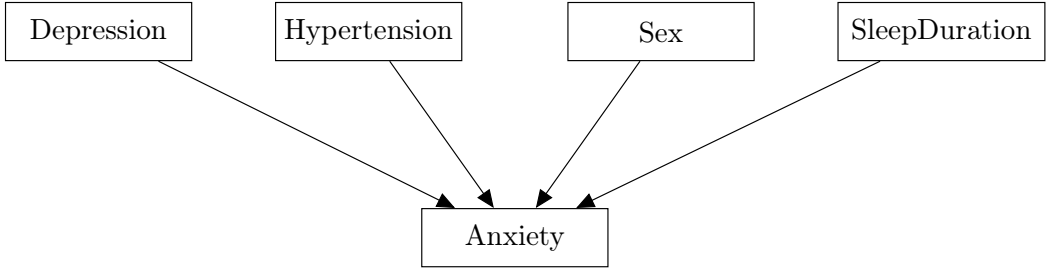
\begin{figure}[h]
    \centering
    \scalebox{1.1}{\begin{tikzpicture}
    \tikzset{mynode/.style={latent, minimum width=2.25cm, shape=rectangle, inner sep=0pt, align=center}}

    \node[mynode] (Y1) {Anxiety};
    \node[above=of Y1, yshift=1cm] (Xmid1) {};
    \node[mynode, left=of Xmid1, xshift=0.5cm] (X21) {Hypertension};
    \node[mynode, left=of X21,xshift=-0cm] (X11) {Depression};
    \node[mynode, right=of Xmid1, xshift=-0.5cm] (X-11) {Sex};
    \node[mynode, right=of X-11,xshift=0cm, minimum width = 2.5cm] (Xn1) {SleepDuration};

    \edge {X11,X21,X-11,Xn1} {Y1};

    \end{tikzpicture}}
    \caption{Local structure of the Anxiety node in the Cardiovascular BN \cite{Ordovas2023}}
    \label{fig:AnxietyNode}
\end{figure}

As CPT approximation methods become more necessary and of greater practical benefit when the number of parameters being approximated grows, we shortlisted nodes featuring at least four parents on which to evaluate these structural methodologies. To ensure clarity and accessibility of our worked example, we focused on binary nodes with exactly four parents. This left us just the nodes `Diabetes' and `Anxiety', and we opted to focus on the latter. The local structure of the Anxiety node is shown in Figure \ref{fig:AnxietyNode}.

The local state structure of the Anxiety node is $(2,2,2,3)\rightarrow2$, yielding a CPT with 24 rows and 24 free parameters (48 total). Without a suitably large dataset, such as the one used to parameterise the true model, the modelling of this CPT through data alone could lead to unreliable and unstable parameter estimates - especially as $\mathbb{P}(\text{Anxiety}=\text{Yes}|\mathbf{X})$ is often low. Henceforth, we suppose that no dataset of sufficient size and quality is available with which to directly model the Anxiety node following the structure shown in Figure \ref{fig:AnxietyNode}.  

Modifying the structure of the Anxiety node may enable reliable, stable parameter estimates to be made if just a limited supply of data is available. However, if this limited data remains insufficient, expert judgement will be required to obtain reliable parameter estimates. If expert judgement is required, even if in combination with a limited dataset, it would be costly and inefficient to formally elicit every parameter of the original CPT. In either case, there is a clear benefit of reducing the parameter space of the Anxiety node through an appropriately chosen structural refinement of its local structure. 

Reducing the parameter space of its CPT does, however, come at a cost of reduced flexibility and faithfulness. Because of this, we will later apply each of the below structural methods to the local structure of the Anxiety node, evaluating the parameter savings of each method as well as the minimum possible information loss each brings in its best-case scenario. We optimise the approximate CPT when using each method over the reduced parameter space it brings, and we compare each approximate CPT to the `true' CPT learnt from data. This process and the subsequent discussion of its output are presented in Section \ref{Comparison}.

\section{Methods} \label{Methods}

\subsection{Edge and Node Pruning}
The most direct way to simplify a local Bayesian network structure is to prune edges within it. Pruning an edge reduces the size of the parent set of the particular child by one, directly reducing the number of parameters in its CPT through an exponential decay (see Equation \ref{CPTParameterEq}). Pruning a node $X_i$ simply deletes that node and all its adjacent edges from the network, reducing the size of any CPT in which $X_i$ was a parent and removing any parameters used to model $X_i$ itself.

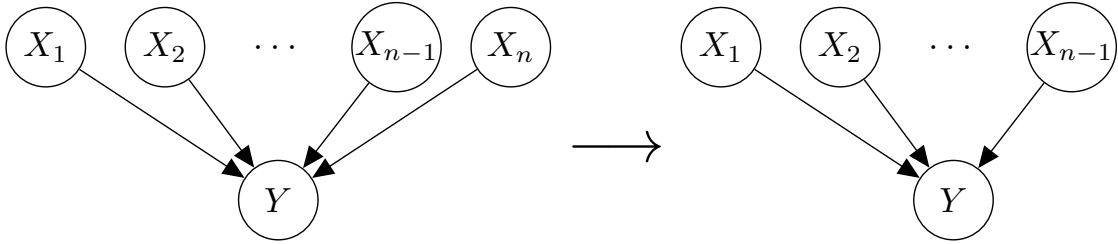
\begin{figure}[h]
    \centering
    \scalebox{1.32}{\begin{tikzpicture}
    \tikzset{mynode/.style={latent, minimum size=0.8cm, shape=circle, inner sep=0pt, align=center}}

    \node[mynode] (Y1) {$Y$};
    \node[above=of Y1] (Xmid1) {};
    \node[mynode, left=of Xmid1, xshift=0.4cm] (X21) {$X_2$};
    \node[mynode, left=of Xmid1,xshift=-0.8cm] (X11) {$X_1$};
    \node[mynode, right=of Xmid1, xshift=-0.4cm] (X-11) {$X_{n-1}$};
    \node[mynode, right=of Xmid1,xshift=0.8cm] (Xn1) {$X_n$};

    \path (X21) -- node[auto=false]{\ldots} (X-11);
    
    \edge {X11,X21,X-11,Xn1} {Y1};

    \begin{scope}[xshift=6.8cm]
        \node[mynode] (Y2) {$Y$};
        \node[above=of Y2] (Xmid2) {};
        \node[mynode, left=of Xmid2, xshift=0.4cm] (X22) {$X_2$};
        \node[mynode, left=of Xmid2,xshift=-0.8cm] (X12) {$X_1$};
        \node[mynode, right=of Xmid2, xshift=-0.4cm] (X-12) {$X_{n-1}$};
    \node[below=of X12,xshift=0.8cm,yshift=0.5cm] (M) {};
    \path (X22) -- node[auto=false]{\ldots} (X-12);
    
    \edge {X12,X22,X-12} {Y2};
    \end{scope}
     \node[draw=none, font=\bfseries\LARGE] at ($(M -| Y1)!0.5!(M -| Y2)$) {$\longrightarrow$};
    \end{tikzpicture}}
    \caption{Before and after pruning the edge $(X_n,Y)$ in a local Bayesian network structure}
    \label{fig:Pruning}
\end{figure}

Suppose we have a child node $Y$ with parent set $\mathbf{X}$ of size $|\mathbf{X}|=n$. We can reduce the parameter space of the CPT $Y|\mathbf{X}$ through pruning an edge, say $(X_p,Y)$. This process, for $p=n$, is demonstrated in Figure \ref{fig:Pruning}. Approximating the full CPT through parameterising the pruned CPT of $Y|(\mathbf{X}\setminus X_p)$ is performed through the simple approximation formula:
\begin{equation}\label{PruningEq}
    p(y|\mathbf{x})=p\left(y\mid\mathbf{x}_{-p}=\mathbf{x}\setminus x_p\right).
\end{equation}

Pruning an edge can provide great parameter savings, but it can also lead to high information loss if the parent being disconnected from the child has a strong influence on the child. We focus therefore on pruning just one node rather than multiple. When pruning the edge $(X_p,Y)$, the number of parameters needing to be defined in the approximate, pruned CPT is given as $N_Y/s_p$. In the $n$-parent case where all nodes are binary, pruning one edge brings a parameter saving of $2^n-2^{n-1}=2^{n-1}$, a saving of 50\%. The approximated CPT of $Y|(\mathbf{X}\setminus X_p)$ can be used to populate the full CPT of $Y|\mathbf{X}$, providing a final approximate CPT following the structure of the original model. Any rows in the full CPT that have a common partial configuration across the parents $\mathbf{X}\setminus X_p$ will have the same CPD defined across the states of the child node $Y$, as defined by the row of the pruned CPT with that configuration.

Edge pruning can be performed through expert judgement or through data-driven approaches. The goal is to remove the edges that correspond to the weakest dependency structures. This corresponds to disconnecting the least influential parent(s) from the child. Domain experts are able to provide judgements regarding the strength of influence of each parent without much difficulty, and this is even required in many quantitative CPT approximation methods \cite[e.g.][]{Hassall2019,Fenton2007,Wisse2008,Roed2009}. These judgements can then be used to determine which edges, if any, can be pruned without significant information loss. Data-driven approaches to edge pruning similarly focus on the goal of minimising information loss when removing edges from the network \cite[e.g][]{BaratySimovici2009,Choi2005}, or address the problem of identifying irrelevant nodes given a particular target node using ideas of d-separation and barren nodes \cite[e.g.][]{BakerBoult1990}. Pruning can be used in either case, though it should generally only be considered in cases where one or more parents have a notably low influence on the child.

\subsection{Divorcing} \label{Divorcing}
Divorcing refers to partitioning a parent set into two blocks, with one block passing through an intermediate node before reaching the child \cite{JensenNielsen2007,KorbNicholson2011}. While this leads to greater structural complexity, divorcing can provide significant parameter savings.

The main goal when divorcing parents is to group similar parents - those whose causal mechanisms overlap or interact the most \cite{Rohrbein2009}. By doing this, the intermediate node can be defined far simpler and more intuitively than if two semantically and mechanistically distant nodes were placed together. Furthermore, it is impossible to model any interactions between divorced parents \cite{KorbNicholson2011}, hence it is important to group parents that have the strongest interactions. The intermediate node combining the divorced parents can often be defined through a simple deterministic operator - as seen by the `Tuberculosis or Lung Cancer' node in the Asia Bayesian network example \cite{LauritzenSpiegelhalter1988} - but it can also be treated stochastically like any other node \cite{Olesen1989}.

\begin{figure}[ht!]
    \centering
    \scalebox{1.17}{\begin{tikzpicture}
    \tikzset{mynode/.style={latent, minimum size=0.8cm, shape=circle, inner sep=0pt, align=center}}

    \node[mynode] (Y1) {$Y$};
    \node[above=of Y1] (Xmid1) {};
    \node[mynode, left=of Xmid1, xshift=0.9cm] (Xi1) {$X_i$};
    \node[mynode, left=of Xi1,xshift=0.2cm] (X11) {$X_1$};
    \node[mynode, right=of Xmid1, xshift=-0.9cm] (Xi1p1) {$X_{i+1}$};
    \node[mynode, right=of Xi1p1,xshift=-0.2cm] (Xn1) {$X_n$};

    \path (X11) -- node[auto=false]{\ldots} (Xi1);
    \path (Xi1p1) -- node[auto=false]{\ldots} (Xn1);

    \edge {X11,Xi1,Xi1p1,Xn1} {Y1};

    \begin{scope}[xshift=7cm]
        \node[mynode] (Y2) {$Y$};
        \node[above=of Y2] (Xmid2) {};
        \node[mynode, left=of Xmid2, xshift=0.9cm] (Xi2) {$X_i$};
        \node[mynode, left=of Xi2,xshift=0.2cm] (X12) {$X_1$};
        \node[mynode, right=of Xmid2, xshift=-0.9cm] (Xi2p1) {$X_{i+1}$};
        \node[mynode, right=of Xi2p1,xshift=-0.2cm] (Xn2) {$X_n$};
        \node[mynode, below=of X12,xshift=0.8cm,yshift=0.9cm] (M) {$M$};
        \path (X12) -- node[auto=false]{\ldots} (Xi2);
        \path (Xi2p1) -- node[auto=false]{\ldots} (Xn2);

        \edge {Xi2p1,Xn2,M} {Y2};
        \edge {X12,Xi2} {M};
    \end{scope}
    \node[draw=none, font=\bfseries\LARGE] at ($(M -| Y1)!0.5!(M -| Y2)$) {$\longrightarrow$};
    \end{tikzpicture}}
    \caption{Before and after divorcing parents in a local Bayesian network structure}
    \label{fig:Divorcing}
\end{figure}
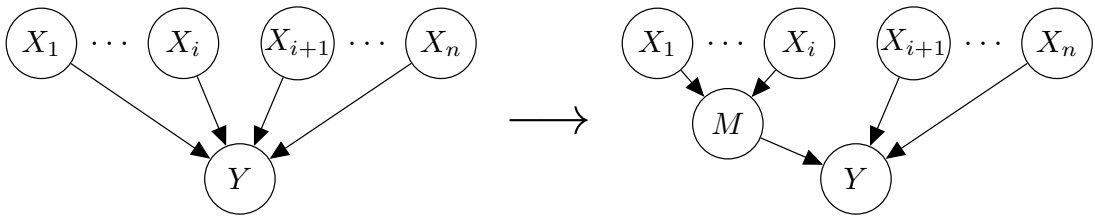

The structure resulting from the divorcing process is demonstrated in Figure \ref{fig:Divorcing} in which we have $n$ parents, the first $i$ of which we divorce through the intermediate node $M$. In this general case, assuming $M$ is defined deterministically, we obtain the following approximation formula with which to approximate the original CPT of $Y|\mathbf{X}$, where $\mathbf{X}_{(i)}=\{X_1,\ldots X_i\}$:
\begin{equation}
    p(y|\mathbf{x})= p\left(y|x_{i+1},\ldots,x_n,m=f(\mathbf{x}_{(i)})\right).
\end{equation}
If $M$ is indeed deterministic, this approximation simply requires the CPT of \\$Y|\{M,X_{i+1},\ldots,X_n\}$ to be determined. If all nodes are binary, this CPT features $2^{n-i+1}$ free parameters, resulting in a parameter saving of $2^n-2^{n-i+1}=2^{n-i+1}(2^{i-1}-1)$ parameters.

In the simplest cases, just two parents are divorced from the rest - in which $i=2$ for the general case shown in Figure \ref{fig:Divorcing}. This simple case allows good freedom to model interactions across the remaining parents yet nonetheless yields an beneficial parameter saving. Divorcing just two parents renders it relatively natural and intuitive to find a deterministic operator with which to define $M$, usually comprising a simple Boolean operator such as AND, OR or XOR. It is equally possible to divorce a greater number of parents to further reduce the parameter space. The warning here is that this may reduce the flexibility and faithfulness of the resulting model, as well as increasing the difficulty associated with defining $M$.

Algorithmic data-driven approaches for choosing suitable parents to divorce are typically based on the general notion of grouping parents by similarity \cite{Rohrbein2009,Rosenkrantz2025}. The use of expert judgement for parent divorcing is less explored, though it seems a natural approach for determining which parents naturally interact the most in the real-world system being modelled, and therefore which parents should be divorced. Defining a divorced model through the use of expert judgement is demonstrated by Case Study 2 in \cite{Boneh2006}. Divorcing is a good approach to take, whether using data or expert judgement, when pruning leads to unsatisfactory information loss and when there is a natural grouping of even just a small number of parents from the rest. The remaining three structural methods build on the general divorcing methodology, utilising intermediate nodes holistically across the entire parent set.

\subsection{Simple Canonical Models} \label{SCMs}
Simple canonical models (SCMs) \cite{DiezDruzdzel2006} form a very basic class of causal interaction model - the first of three that we evaluate in this paper. A causal interaction model introduces a layer of independent mechanism nodes between the child and its parents \cite{MeekHeckerman1997}, and can be seen as an extension to the general divorcing methodology.

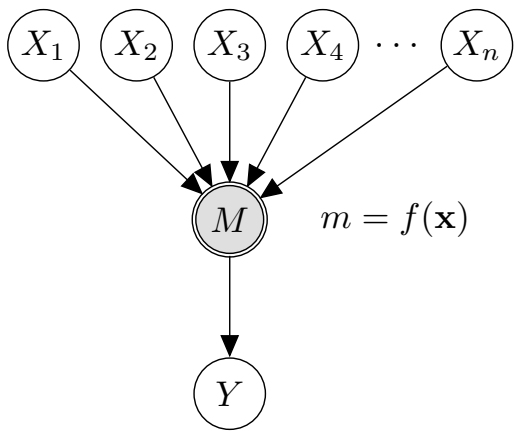
\begin{figure}[ht]
\centering
\scalebox{1.35}{
\begin{tikzpicture}
     \node[latent,xshift=-0.8cm] (Y) {$Y$};%
     \node[obs,above=of Y,style = double] (M) {$M$}; %
     \node[latent,above=of M] (X3) {$X_3$};
     \node[latent,left=of X3,xshift=0.8cm] (X2) {$X_2$};
     \node[latent,left=of X2,xshift=0.8cm] (X1) {$X_1$};
     \node[latent,right=of X3,xshift=-0.8cm] (X4) {$X_4$};
     \node[latent,right=of X4,xshift=-0.2cm] (Xn) {$X_n$};
     \node[right=of M,yshift=0cm,xshift=-0.6cm] (f){\small{$m=f(\mathbf{x})$}};;
     \path (X4) -- node[auto=false]{\ldots} (Xn);
    \coordinate[right of=X4,xshift=-0.35cm,yshift=-0.2cm] (d1);
    \coordinate[right of=d1,xshift=-0.4cm] (d2);
     \edge {M} {Y};
     \edge {X1,X2,X3,X4,Xn} {M};
\end{tikzpicture}
}
\caption{The general structure of a simple canonical model \cite{DiezDruzdzel2006}}
\label{fig:SCM}
\end{figure}

In the SCM structure, the intermediate layer comprises just one node, denoted $M$, which each parent directly connects to. The child node $Y$ now just has the one parent, $M$, whereas the new intermediate node has parent set $\mathbf{X}$ of size $n$. The structure of an SCM is shown in Figure \ref{fig:SCM}.

As indicated in Figure \ref{fig:SCM}, the intermediate node $M$ is modelled deterministically in the SCM framework. The only relationship that is modelled stochastically is that of $Y|M=f(\mathbf{X})$. Therefore, we use the following, simple formula to approximate the original CPT of $Y|\mathbf{X}$ when assuming the SCM structure:
\begin{equation}\label{SCMeq}
    p(y|\mathbf{x})=p(y|m=f(\mathbf{x})).
\end{equation}
This produces the most extreme parameter saving of any structural method in this paper. In the simplest case that $Y$ and $M$ are both binary, there are just two parameters to determine. Increasing the number of parents, or the number of states per parent, does not increase this number of required parameters, but it can make the definition of $f(\mathbf{X})$ much harder. In the case that each node in the structure is binary, it is clear to see that an SCM brings a parameter saving of $2^n-2$ parameters.

An example of an SCM over binary variables is the `simple AND' model \cite{DiezDruzdzel2006} in which the deterministic combination function $f$ is the AND function over the parent set $\mathbf{X}$. The relationship $Y|M$ is characterised by $\mathbb{P}(Y=1|M=1)=c$, representing the probability that the effect is indeed present when the necessary causes are present, and $\mathbb{P}(Y=1|M=0)=s$, representing the probability that - despite lacking the necessary causes - the effect in the child is seen nonetheless. These two parameters are sufficient for parameterising the entire SCM.

Defining an SCM relies on the ability to elicit a suitable deterministic combination function $f$ with which to model the intermediate node $M$. This function effectively partitions the set of configurations $\mathbf{x}$ of the parent nodes into blocks that correspond to each of the child's states, with each block being assigned a common CPD across the states of the child node. Especially when $n$ grows large, it can be extremely challenging, if not impossible, to find a satisfactory combination function that is faithful to expert beliefs about the real-world system. However, if the real-world system features some largely deterministic components, an SCM could be a very efficient way to represent it without much information loss. The SCM framework is, as a natural consequence of providing such extreme parameter savings, the least flexible structural method we present, and hence is the least applicable to real-world modelling projects.


\subsection{Independence of Causal Influences} \label{ICI}

\begin{figure}[ht]
\centering
\scalebox{1.35}{
\begin{tikzpicture}
     \node[latent,xshift=-1.2cm,style = double] (Y) {$Y$};%
     \node[obs,above=of Y,xshift=-1.8cm] (M1) {$M_{1}$}; %
     \node[obs,right=of M1,xshift=-0.8cm] (M2) {$M_{2}$}; %
     \node[obs,right=of M2,xshift=-0.8cm] (M3) {$M_{3}$}; %
     \node[obs,right=of M3,xshift=-0.8cm] (M4) {$M_{4}$}; %
     \node[obs,right=of M4,xshift=0.2cm] (Mn) {$M_n$}; %
     \node[latent,above=of M1] (X1) {$X_1$};
     \node[latent,right=of X1,xshift=-0.8cm] (X2) {$X_2$};
     \node[latent,right=of X2,xshift=-0.8cm] (X3) {$X_3$};
     \node[latent,right=of X3,xshift=-0.8cm] (X4) {$X_4$};
     \node[latent,right=of X4,xshift=0.2cm] (Xn) {$X_n$};
     \node[right=of Y,yshift=0cm,xshift=-0.6cm] (f){\small{$y=f(\mathbf{m})$}};;
     \path (M4) -- node[auto=false]{\ldots} (Mn);
     \path (X4) -- node[auto=false]{\ldots} (Xn);
    \coordinate[right of=X4,xshift=-0.35cm,yshift=-0.2cm] (d1);
    \coordinate[right of=d1,xshift=-0.4cm] (d2);
     \edge {M1,M2,M3,M4,Mn} {Y};
     \edge {X1} {M1};
     \edge {X2} {M2};
     \edge {X3} {M3};
     \edge {X4} {M4};
     \edge {Xn} {Mn};
\end{tikzpicture}
}
\caption{The structure of the ICI model}
\label{fig:ICIstructure}
\end{figure}
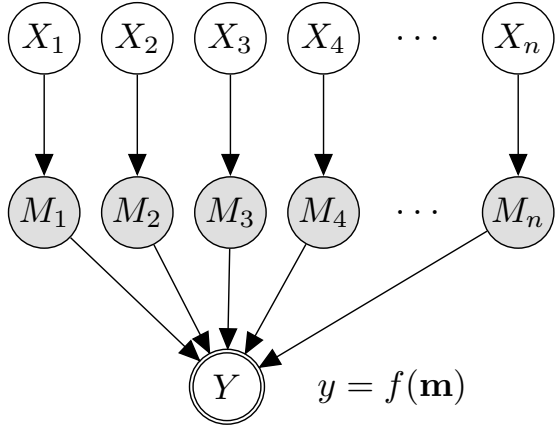

A more expressive class of causal interaction model is that of the \textit{independence of causal influences} (ICI) model \cite{Heckerman1993}, historically also referred to as causal independence models. The ICI model, as a causal interaction model, introduces a layer of mechanism nodes between the child and its parent set. Unlike SCMs, this layer of mechanisms comprises multiple nodes. In the ICI model, each parent, $X_i$, connects directly to exactly one intermediate mechanism node, $M_i$, that is unique to that parent. This defines a bijection between the parent set $\mathbf{X}$ and the mechanism set $\mathbf{M}$, denoting this mapping by $\phi:\mathbf{X}\rightarrow\mathbf{M}$. The structure of the ICI model is illustrated in Figure \ref{fig:ICIstructure} \cite{Heckerman1993}.

In the ICI model, the mechanism nodes are defined stochastically, while the child node is modelled through the deterministic function $f$ over the set of mechanism nodes, as indicated in Figure \ref{fig:ICIstructure}. The CPD for a given row in the true CPT of $Y|\mathbf{X}$ can be approximated through the following probability mass function that defines the ICI model \cite{vanGerven2008}:
\begin{align}\label{ICIdef}
p(y|\mathbf{x})=\sum_{\mathbf{m}|f(\mathbf{m})=y}\;\prod_{i=1}^{n}p(m_{i}|x_i).
\end{align}

The ICI model provides significant parameter savings coming from the assumption that the mechanisms operate independently, and through the use of a deterministic combination function to model the combined effects of these mechanisms on the child. The number of quantitative parameters required to define an ICI model, assuming the child and hence the intermediate mechanism nodes to be binary, is the sum of the number of states of each parent, written as $s_1+s_2+\ldots+s_n$. This yields a parameter saving of $\prod_{i=1}^ns_i-\sum_{i=1}^ns_i$ for a binary child. Some specific subclasses of ICI model, such as the noisy OR model \cite{Pearl1988ProbReasoning} and its extensions \cite{DiezDruzdzel2006}, actually require even fewer parameters to be determined because they impose further quantitative constraints on the ICI parameters. In particular, some parameters are assumed to be zero, as can be seen in the noisy OR example below.

\begin{figure}[ht]
    \centering
    \scalebox{1.01}{
    \begin{tikzpicture}
     \node[latent,style = double] (Y) {$Y$};%
     \node[obs,above=of Y,xshift=-2.8cm] (M1) {$M_{1}$}; %
     \node[obs,above=of Y,xshift=0cm] (M2) {$M_{2}$}; %
     \node[obs,above=of Y,xshift=2.8cm] (Mm) {$M_{3}$}; %
     \node[latent,above=of M1] (X1) {$X_1$};
     \node[latent,above=of M2] (X4) {$X_2$};
     \node[latent,above=of Mm] (Xn) {$X_3$};
     \node[below=of X4,yshift=-1.3cm,xshift=-4.25cm] (f1){
\scalebox{0.8}{
\begin{tabular}{|c|c|c|}
\hline
$X_i$ & $\mathbb{P}(M_i=0\mid X_i)$ & $\mathbb{P}(M_i=1\mid X_i)$\\
\hline
$0$   &  $1$    &   $0$    \\
$1$   &  $p_i$    &   $1-p_i$    \\
\hline  
\end{tabular}
}
};
    \coordinate[right of=X4,xshift=-0.35cm,yshift=-0.2cm] (d1);
    \coordinate[right of=d1,xshift=-0.4cm] (d2);
     \edge {M1,M2,Mm} {Y};
     \edge {X1} {M1};
     \edge{X4} {M2};
     \edge {Xn}{Mm};
     \node[right=of Y,yshift=0cm,xshift=-1cm] (f){\small{$y=f(\mathbf{m})=\bigvee\limits_{i=1}^{3}M_i$}};
\end{tikzpicture}
}
\caption{The noisy OR model as an explicit ICI model}
\label{fig:ICINoisyOR}
\end{figure}
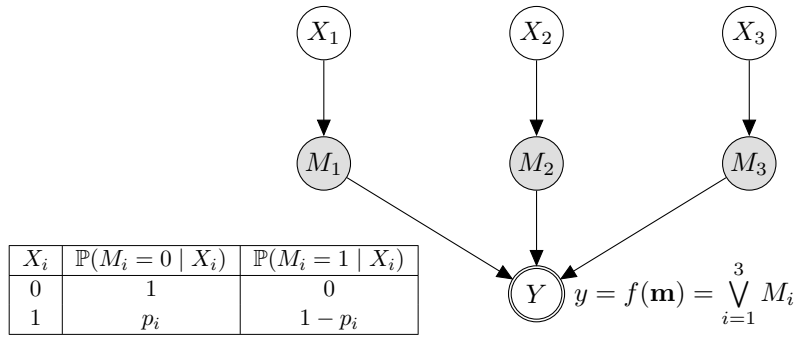

In the noisy OR model, which is defined over a set of binary variables, the mechanism node of each parent has the ability to inhibit the causal state of its parent when it is observed (i.e. $\mathbb{P}(M_i=0\mid X_i=1)\geq0$), but not to enforce the effect of that causal state to the child if it is not observed (i.e. $\mathbb{P}(M_i=0|X_i=0)=1$). Each mechanism node, $M_i$, is therefore simply parameterised by $\mathbb{P}(M_i=0|X_i=1)=p_i$. This is an example of an ICI model that is fully embellished with just $n<2^n$ parameters. The noisy OR model construction explicitly as an ICI model is shown in Figure \ref{fig:ICINoisyOR} \cite{Drury2025SICI}.

One particular generalisation of the ICI model, known as the \textit{probabilistic independence of causal influences} (PICI) model, allows the relationship $Y|\mathbf{M}$ to also be modelled stochastically \cite{ZagoreckiDruzdzel2006,DiezDruzdzel2006,Zagorecki2006}. The structure of the PICI model is equal to that of the ICI model, and is shown in Figure \ref{fig:ICIstructure} (except for the child node being modelled deterministically). The PICI model represents $\mathbf{M}|\mathbf{X}$ stochastically, as does the standard ICI model, but it also demands a stochastic relationship for $Y|\mathbf{M}$. The probability mass function $p(y|\mathbf{x})$ defining the PICI model is thus \cite{DiezDruzdzel2006}:
\begin{align}
    p(y|\mathbf{x})=\sum_\mathbf{m}\left[p(y|\mathbf{m})\prod_{i=1}^np(m_i|x_i)\right].\label{eq:PICI}
\end{align}

The now stochastic representation of $Y|\mathbf{M}$ is generally captured by a CPT with parent set $\mathbf{M}$ of size $n$. This CPT often features just as many (or nearly as many) parameters as the original CPT of $Y|\mathbf{X}$. In addition to the $s_1+\ldots+s_n$ parameters needed to model $\mathbf{M}|\mathbf{X}$, parameter savings through the use of the PICI model without any further quantitative restrictions are minimal, if at all possible.

The noisy average model \cite{Zagorecki2006,ZagoreckiDruzdzel2006,DiezDruzdzel2006} is an example of a PICI model that does implement an additional quantitative restriction on the modelling of $Y|\mathbf{M}$. In this model, the mechanism nodes are defined to have the same state space as the child node, and the probability mass function $p(y|\mathbf{m})$ is defined by the following averaging function:
\begin{equation*}
    f(y,\mathbf{m})=p(y|\mathbf{m})=\frac{1}{n}\left|\{m_i|m_i=y\} \right|=\frac{1}{n}\sum_{1=1}^n\mathds{1}_{\left\{m_i=y\right\}}
\end{equation*}
While this model features the same number of parameters as the ICI model, the process of eliciting a stochastic function with which to model $Y|\mathbf{M}$ is potentially more complex than eliciting a deterministic function for this relationship. Allowing a stochastic relationship here does introduce additional flexibility over the standard ICI model, but it would be difficult to efficiently elicit such a relationship while maintaining faithfulness to expert beliefs about the real-world system. For our worked example, we evaluate the performance of the standard ICI model as the parameter savings associated with the PICI model are minimal without further quantitative restrictions that may be very difficult to elicit.

\subsection{Surjective Independence of Causal Influences} \label{SICI}

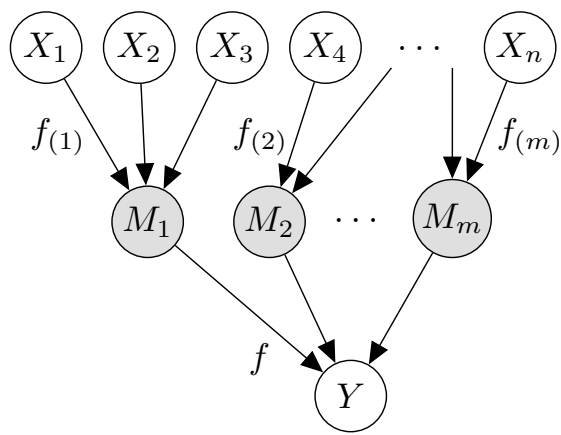
\begin{figure}[ht]
\centering
\scalebox{1.35}{
\begin{tikzpicture}
     \node[latent] (Y) {$Y$};%
     \node[obs,above=of Y,xshift=-2cm] (M1) {$M_{1}$}; %
     \node[obs,above=of Y,xshift=-0.8cm] (M2) {$M_{2}$}; %
     \node[obs,above=of Y,xshift=1cm] (Mm) {$M_{m}$}; %
     \node[latent,above=of M1,xshift=-1cm] (X1) {$X_1$};
     \node[latent,right=of X1,xshift=-0.8cm] (X2) {$X_2$};
     \node[latent,right=of X2,xshift=-0.8cm] (X3) {$X_3$};
     \node[latent,right=of X3,xshift=-0.8cm] (X4) {$X_4$};
     \node[latent,right=of X4,xshift=0.2cm] (Xn) {$X_n$};
     \node[below=of X2,yshift=0.8cm,xshift=-0.8cm] (f1){\small{$f_{(1)}$}};
     \node[below=of X4,yshift=0.8cm,xshift=-0.64cm] (f2){\small{$f_{(2)}$}};
     \node[below=of Xn,yshift=0.8cm,xshift=0.1cm] (fm){\small{$f_{(m)}$}};
      \node[below=of M1,yshift=0.3cm,xshift=1.1cm] (f){\small{$f$}};
     \path (M2) -- node[auto=false]{\ldots} (Mm);
     \path (X4) -- node[auto=false]{\ldots} (Xn);
    \coordinate[right of=X4,xshift=-0.35cm,yshift=-0.2cm] (d1);
    \coordinate[right of=d1,xshift=-0.4cm] (d2);
     \edge {M1,M2,Mm} {Y};
     \edge {X1,X2,X3} {M1};
     \edge{X4,d1} {M2};
     
     \edge {Xn,d2}{Mm};
\end{tikzpicture}
}
\caption{The general SICI model structure for a particular partition $\phi$ of the parent set}
\label{SICIstructure}
\end{figure}

A recent generalisation of the ICI methodology is the class of surjective independence of causal influences (SICI) models \cite{Drury2025SICI}. The SICI model, also being a causal interaction model, introduces a layer of intermediate mechanism nodes, but embeds a surjective mapping $\phi:\mathbf{X}\rightarrow\mathbf{M}$ between the parent nodes and the mechanism nodes. This weakens the bijective assumption in the ICI model. The SICI model thereby allows multiple parents to share a common causal mechanism, thus allowing interactions between parents - though these causal mechanisms are still assumed to operate independently. As a result, the SICI model features $m\leq n$ intermediate mechanism nodes, and this structure is shown in Figure \ref{SICIstructure} \cite{Drury2025SICI}.

A key goal of modelling with the SICI model is to embed the assumption of ICI across the set of mechanisms $\mathbf{M}$ through the choice of surjection $\phi$. This is generally performed by grouping parents based on the strength of their interactions, ensuring that parents that lead to different mechanism nodes only have weak interactions, if any. Therefore, this surjective mapping can be determined through partitioning the parent set into blocks of parents who share highly interdependent causal mechanisms with respect to the particular child node. These blocks act as categorisations of the parent nodes, and it is more reasonable to assume the ICI property to hold across these blocks than it is to assume this as a property of the original parent set. This then justifies the incorporated use, if required, of quantitative CPT approximation techniques that rely on, or otherwise benefit from, the ICI property \cite{Drury2025SICI}. 

There are three particular variants of SICI model, each sharing the same structure as shown in Figure \ref{SICIstructure}. The three variants of the SICI model provide three different approaches for parameterising the model, leading to different formulae for $p(y|\mathbf{x})$ with which we approximate the CPT of $Y|\mathbf{X}$ \cite{Drury2025SICI}.
The most general variant of SICI model is the double-stochastic SICI model (DS-SICI) in which every node in the model is modelled stochastically. Another variant is the upper-stochastic SICI model (US-SICI) in which only the upper relationships defining $\mathbf{M}|\mathbf{X}$ are modelled stochastically. The remaining variant is the lower-stochastic SICI model (LS-SICI) in which only the lower relationship defining $Y|\mathbf{M}$ is stochastic. Discussion about the benefits and limitations of each variant, as well as formulae for calculating the approximated CPT of $Y|\mathbf{X}$, can be found in \cite{Drury2025SICI}. Here, we provide the probability mass function defining the DS-SICI model, the most general of the SICI models, from which the formula for the US-SICI and LS-SICI models can be obtained \cite{Drury2025SICI}, noting that $\mathbf{X}_{(i)}=\{X_j:\phi(X_j)=M_i\}$:
\begin{equation}\label{eq:DS-SICI}
    p(y|\mathbf{x})=\sum_{\mathbf{m}}p(y|\mathbf{m},\mathbf{x})p(\mathbf{m}|\mathbf{x})=\sum_{\mathbf{m}}\left(p(y|\mathbf{m})\prod_{i=1}^mp(m_i|\mathbf{x}_{(i)})\right).
\end{equation}

In this paper, we evaluate the performance of the US-SICI model as it is the closest in nature to the standard ICI model, allowing a meaningful comparison between the two methods. The probability mass function defining the US-SICI model is given as \cite{Drury2025SICI}:
\begin{equation}\label{eq:US-SICI}
    p(y|\mathbf{x})=\sum_{\mathbf{m}}p(y|\mathbf{m},\mathbf{x})p(\mathbf{m}|\mathbf{x})=\sum_{\mathbf{m}\mid f(\mathbf{m})=y}\prod_{i=1}^mp(m_i|\mathbf{x}_{(i)}).
\end{equation}
The quantitative parameters defining the US-SICI model are only present in the upper relationships of the model. As such, the number of parameters the model demands is equal to the number of free parameters across the CPTs of $M_i|\mathbf{X}_{(i)}$. In the fully binary setting, this total number of parameters is calculated as $2^{\left|\mathbf{X}_{(1)}\right|}+\ldots+2^{\left|\mathbf{X}_{(m)}\right|}$. Where $m=1$, this gives rise to $2^n$ parameters, whereas $m=n$ (i.e. the ICI model) yields $2n$ parameters. All other cases require a quantity of parameters between these bounds.

Two examples of SICI models are presented in the introductory paper to SICI \cite{Drury2025SICI}. The first is the surjective noisy OR model - an example of a US-SICI model. This model is similar to the standard noisy OR model \cite{Pearl1988ProbReasoning}, except multiple parent nodes combine into a mechanism node before the combined causal effect of the parents may be inhibited. The second example applies to either DS-SICI or LS-SICI models as it is a demonstration of how a CPT interpolation algorithm - Hassall's algorithm \cite{Hassall2019} - that implicitly makes use of the ICI assumption can be used to parameterise a CPT within the SICI framework. Details of these examples are omitted here for brevity but can be followed in the original paper \cite{Drury2025SICI}. 

The SICI model provides the most structurally complex methodology, and often yields lower parameter savings than other structural methods. Nevertheless, it should generally be considered an option when dealing with a parent set in which many interacting effects are present, as we will discuss later.

\section{Results and Comparison} \label{Comparison}
We proceed to evaluate the above methods by applying them to the Anxiety node in the Cardiovascular BN \cite{Ordovas2023} alluded to in Section \ref{Example}. For each method, we apply the appropriate structural refinement to the local structure of the Anxiety node and optimise the approximate CPT parameters such that we minimise the sum of total variation distances when comparing the distributions of each CPT row-by-row.

\subsection{Total variation distance and optimisation}
We have a choice of measures with which we can quantify the quality of the CPT approximation. One option that is often used within AI and machine learning is the Kullback-Leibler divergence which has been used previously in the context of evaluating CPT approximations \cite{Blomaard2025}. While our methodology would easily utilise Kullback-Leibler divergence as an evaluation tool, we opt for a similarity measure that is less sensitive at the tails due to the difficulty in accurately eliciting extreme probabilities through expert judgement \cite{ohaganbook}, and as many of the true CPT parameters are probabilities close to 0 or 1. The total variation distance is an intuitive alternative measure that is less sensitive at the tails, hence we choose to use it over the Kullback-Leibler divergence to judge the quality of the optimal approximate CPT generated through each structural methodology.

As we are evaluating CPTs in which the child is discrete and binary, the total variation distance as used here can be reduced in the following way (where $\mathcal{Y}$ denotes the support of the child node $Y$):
\begin{align}
D_{TV}(P,Q) 
&= \frac{1}{2} \sum_{y\in\mathcal{Y} }\left| P(y) - Q(y) \right|=\frac{1}{2} \left( \left| P(y_0) - Q(y_0) \right| + \left| P(y_1) - Q(y_1) \right| \right) \hspace{2cm}\notag \\
&= \frac{1}{2} \left( \left| P(y_0) - Q(y_0) \right| + \left| 1 - P(y_0) - \bigl(1 - Q(y_0)\bigr) \right| \right) \notag \\
&= \frac{1}{2} \left( \left| P(y_0) - Q(y_0) \right| + \left| Q(y_0) - P(y_0) \right| \right) \notag \\
&= \left| P(y_0) - Q(y_0) \right|.
\end{align}
We will refer to the true CPD over the two child states for a given row $j$ as $P^j=(p_1^j,p_2^j)$, and the approximate CPD for row $j$ as $Q^j=(q_1^j,q_2^j)$. We may also refer to the distribution $Q^k=(q_1^k,q_2^k)$ to indicate the $k^\text{th}$ unique distribution in the approximate CPT. Row indices may be omitted where the rows in focus are clear, or where a general row is being discussed.

In some of the methods below, the structural refinement imposed leads to groupings $g_k$ of rows in the full CPT that are each forced to share a common distribution over the child states. One approximate distribution $Q^k=(q_1^k,q_2^k)$ will be used to approximate multiple rows, determined by $g_k$, of the full CPT. We denote the size of the grouping $k$ by $s_k=|g_k|$. When taking the sum of the total variation distances between the CPTs row-by-row, the distribution $Q^k$ will contribute a term of $|p_1^{k1}-q_1^k|+\ldots+|p_1^{ks_k}-q_1^k|$, in which $p_1^{k1},\ldots,p_1^{ks_k}$ correspond to the free parameters across the true CPT rows in the grouping $g_k$. As each row in the true CPT belongs to exactly one grouping, we can optimise the approximate CPT parameters through choosing each $q_i^k$ to minimise this term. The optimal parameters are calculated as below, as the solution to the least absolute deviation problem in one-dimension:

\begin{align}
q_1^k=\arg\min_{[0,1]}\sum_{j=1}^{s_k}|p_1^{kj}-q_1|=\text{median}\{p_i^{kj}:j=1,\ldots,s_k\}
\end{align}

For the latter methods that we evaluate, the parameter space reduction does not simply correspond to grouping rows of the true CPT in this way. Instead, smaller CPTs are defined across a set of intermediate nodes that then combine through some combination function to produce an effect on the child, giving rise to particular approximation formulae that remain unique to each parent configuration $\mathbf{x}$. In this case, there is no closed-form solution for obtaining the necessary optimal parameters. In order to optimise these parameters, we can instead utilise real-valued genetic algorithms with a loss function that corresponds to the sum of row-wise total variation distances between the true CPT and the resulting approximate CPT. These genetic algorithms are also able to help optimise over a choice of structures and over a choice of deterministic combination functions where a methodology requires this. Details of this will be provided in the relevant sections below. In general, we ran each genetic algorithm with a population size of 300 candidate solutions, a maximum of 2000 generations, a mutation probability of 0.3, an elitism parameter of 0.05, a crossover probability of 0.8, and a stopping rule of 50 generations without improvement of the best-scoring candidate. This was performed with R v4.5.1 \cite{R4.5.1} through the GA package \cite{GAScrucca2013}.

\subsection{Pruning}
There are four parents that can be pruned from the local structure of the Anxiety node. In order to prevent significant information loss, we focused on pruning just one parent, but multiple parents can be pruned in practice if appropriate. For each parent, we constructed a pruned CPT featuring each configuration of the remaining parents. Each row in this pruned CPT corresponds to $s_p$ rows of the true CPT - where $s_p$ is the number of levels of the parent being pruned. We then optimised each parameter in the pruned CPT through taking the median of the true parameters across the corresponding grouping $g_k$. We scored the optimal CPT approximation for each parent using the sum of row-wise total variation distances, reflecting how similar a CPT it is possible to construct when you prune that parent. Note that this score is a best-case scenario - corresponding to learning or eliciting exactly these optimal parameters which would not be likely in practice.

For example, suppose we prune `Depression' and are evaluating the row of the approximate CPT defined by `Hypertension = Yes', `Sex = Male' and `SleepDuration > 9 hours'. There are two rows in the true CPT that correspond to this partial configuration (formed by adding `Depression = Yes' and `Depression = No' respectively). To compute the parameter representing $\mathbb{P}(\text{Anxiety = Yes}\mid\mathbf{X})$ for this row in the approximate CPT, we take the median of the two corresponding parameters in the true CPT. We do this for every row in the pruned CPT. We duplicate the full CPT structure (i.e. including `Depression' as a parent) and input each approximate parameter where the respective partial configuration of the remaining parents is seen.

The best scoring approximate CPT obtainable by pruning one parent corresponded to pruning the Depression node. The parameters obtained, presented in the same structure in the true CPT, are shown in Table \ref{tab:CPTResults}. This gives an approximate CPT that scores 0.6487 (4dp) by sum of row-wise total variation distances. We also take note that the pruned CPT requires the learning or elicitation of 12 free parameters - down from 24 for the full CPT.

\subsection{Divorcing}
Given the Anxiety node has four parents, we focused on divorcing two parents with which to create an intermediate node. It is possible to divorce just one parent from the rest, or three parents from the last parent, but we consider divorcing two parents from the other two to be the most intuitive approach, providing a good balance between approximation flexibility and parameter savings. By divorcing two parents, we can define the intermediate node through a simple logic gate, resulting in an approximate CPT with three parents (the logic gate and the two remaining parents). Each row in this approximate CPT again corresponds to multiple rows of the true CPT. Two rows of the true CPT that are in grouping $g_k$ must share the same logic gate output and the same values over the remaining two parents (as per row $k$ in the approximate CPT). Again, row $k$ of the approximate CPT is parameterised through taking medians over the parameter sets across the relevant rows of the true CPT. We produced a full approximate CPT by replacing the distributions $\mathbf{P}$ in the true CPT by the approximate distributions $\mathbf{Q}$, ensuring row $j$ featured distribution $Q^k$ if and only if row $j$ was in grouping $g_k$. We then calculated the sum of row-wise total variation distances for each choice of parents being divorced and for each choice of logic gate with which to define the intermediate node.

For example, suppose we divorced `Hypertension' and `SleepDuration' from the remaining parents through an AND gate. Note here that `SleepDuration' is not a binary variable. We handle this simply by mapping one or two of its states to the binary $1$ input, just as we choose one state of `Hypertension' to act as the binary $1$ input. This process thus requires further discretisation into just two states. In this example, any row for which `Hypertension = Yes' and `SleepDuration > 9 hours' will feature a logic gate value of $1$. All rows that share this logic gate output and the same partial configuration over the other two parents will share a common distribution in the approximate CPT.

For each subset of two parents (the divorced parents), we obtained a score when using an AND gate, an OR gate and an XOR gate to define the new intermediate variable. This score utilises the optimal parameterisation through the median function as before. The lowest scoring CPT - hence the best approximation - was found by defining the intermediate node through an AND gate between `Hypertension = Yes' and `SleepDuration > 9 hours', as featured in our example. The parameters obtained are presented in Table \ref{tab:CPTResults}, and produced a score of 0.5072 (4dp), optimised over 8 parameters. There were three groups of size one (coming from the three rows in which the AND gate output was $1$), hence three of the approximate CPT rows featured distributions equal to those in the corresponding rows of the true CPT. This may partially explain why this method performed well, though the remaining parameters each had to be aggregated over a larger number of rows which slightly settles this concern. 

\subsection{SCMs}

The structure of an SCM \cite{DiezDruzdzel2006} is fixed and does not feature any choice of which parents to modify, unlike the above methods. All parents deterministically combine into one intermediate node, $M$, which becomes the sole source of information for $Y$. As Anxiety is a binary variable, we define $M$ to be binary. This setup requires the optimisation of just two approximate CPT parameters (defining $Y|M$). We do, however, have to optimise over the possible functions $f$ that deterministically model $M|\mathbf{X}$. A deterministic function $f$ effectively partitions the parent configurations $\mathbf{x}$ (i.e. the CPT rows) into two groups - one which corresponds to $M=1$, and the other to $M=0$. Such a deterministic combination function would usually be defined through compositions of logic gates. A simple example would define $f$ as an OR gate over all the parents, leading to a mapping for which $M=0$ if and only if the parent configuration $\mathbf{x}$ features the non-causal state of every parent (and $M=1$ otherwise).

Given the state structure of the Anxiety node, there are $2\cdot2\cdot2\cdot3=24$ parent configurations. The partition that defines $f$ must be into exactly two blocks, not necessarily of equal size. We discard any partitions for which the first block is of size greater than $12$ to avoid duplication, leaving the number of non-trivial partitions over which we need to optimise as:
\begin{align}
    \sum_{i=1}^{12} {24\choose i} = \frac{2^{24}}{2}-1=2^{23}-1=8,388,607
\end{align}
With such a large number of partitions to explore, we coded the optimisation problem as an integer-based genetic algorithm.

A given partition forms two groupings - $g_1$ and $g_2$ - of the CPT rows, corresponding to the two rows of the CPT of $Y|M$. Therefore, we optimise over the two free quantitative parameters following the same median method as before. This allows the construction of an optimal approximate CPT for a given partition that can be scored using the total variation distance to form the basis of the loss function for the genetic algorithm.

The optimal approximate CPT that was found through this method featured a partition with size split 8-16, yielding a score of 1.2693 (4dp) by sum of row-wise total variation distances. The parameters of this optimal approximation are presented in Table \ref{tab:CPTResults}. While this would require just two probabilities to be learnt or elicited, there may be a significant struggle to elicit or learn a satisfactory combination function $f$ in practice. The resulting model in this case is not very flexible, and performs relatively poorly.

\subsection{ICI}
An independence of causal influences model also has a fixed structure with one deterministic combination function. This means that, similar to SCMs, we must optimise not only over the quantitative parameters forming the approximate CPTs, but over the combination function $f$. Here, $f$ models the relationship $Y|\mathbf{M}$. Assuming each of the intermediate mechanism nodes to be binary, and given that there is one intermediate node, $M_i$, for each parent, $X_i$, there are $2^4=16$ configurations of mechanism values $\mathbf{m}$. We need to determine the optimal partition of these configurations to optimally define $Y|\mathbf{M}$, again disregarding half of these partitions to avoid duplicate candidate solutions. There are $2^{15}-1$ such non-trivial partitions to optimise over. For this, we utilise a genetic algorithm to search this space as before.

Given a partition of the configurations $\mathbf{m}$, we can then optimise the quantitative parameters of the approximate model. The ICI model demands that each mechanism node is modelled stochastically, with each mechanism $M_i$ requiring one parameter (in this binary setting) per parent state $x_i$. In our case, given the state structure of the Anxiety node, the ICI model features a total of $2+2+2+3=9$ quantitative parameters to be determined. These parameters do not correspond to particular groupings of CPT rows; each row in the approximate CPT features a distinct distribution determined through the formula presented in Section \ref{ICI}. Therefore, there is no closed-form solution for the optimisation of these parameters, and we add these parameters as decision variables in the genetic algorithm used.

We optimise both the combination function $f$ and the set of approximate CPT parameters in the same genetic algorithm. We can no longer run an integer-based genetic algorithm as we have decision variables in the interval $[0,1]$. We can, as part of a mixed-variable genetic algorithm, decode the real-valued results back into integer format as necessary, enabling this joint optimisation to proceed.

The optimal ICI model CPT found by the genetic algorithm resulted in a score of 0.5520 (4dp) with a saving of 15 parameters. The parameters of the optimal approximate CPT are presented in Table \ref{tab:CPTResults}.

\subsection{SICI}
As a generalisation of the ICI model, the SICI model is optimised in much the same way as above. For a given partition of the parent set into $m\leq n$ blocks, the process of optimising the partition of configurations of the mechanism nodes and the quantitative parameters themselves is almost exactly as before. The difference is that the space of partitions of the configurations $\mathbf{m}$ is typically smaller for the SICI model, though this brings a greater number of quantitative parameters than the ICI model. This is due to the introduction of fewer, larger CPTs in the modelling of $\mathbf{M}|\mathbf{X}$. The reduced space of partitions is not nearly substantial enough to use brute-force optimisation methods, and the increased number of parameters over which we optimise is not problematic for the genetic algorithm set-up that we use. Therefore, the genetic algorithm we used for the optimisation of the ICI model is also used here. 

There are a total of $2^n$ partitions of a parent set of size $n$. One of these partitions is the partition of the parents into their own singleton blocks, corresponding to the ICI model. At the other extreme is the partition of the parents into one block that feeds into the sole intermediate node $M$ - a case that does not provide any parameter savings. For the remaining $2^n-2$ partitions, for small to moderate sized $n$, we can perform a brute-force search of the space. This involves modifying the space of decision variables of the genetic algorithm to match the SICI structure that the partition imposes before running the optimisation for each partition. For larger $n$, it would be necessary to introduce more decision variables to the genetic algorithm to include the choice of partition into the search space. This would not be problematic, though it may be necessary to tweak the input parameters to account for the larger search space.

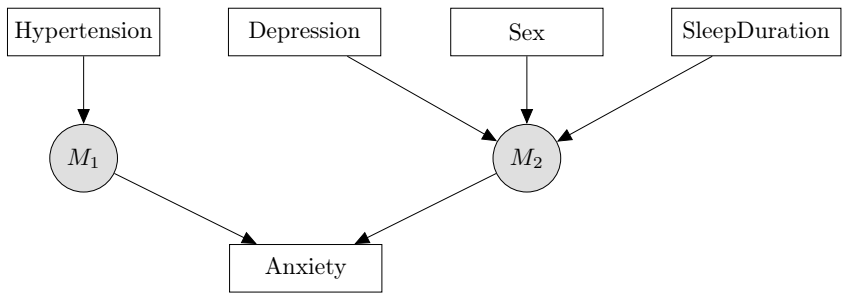
\begin{figure}[h]
    \centering
    \scalebox{0.9}{\begin{tikzpicture}
    \tikzset{mynode/.style={latent, minimum width=2.25cm, shape=rectangle, inner sep=0pt, align=center}}

    \node[mynode] (Y1) {Anxiety};
    \node[above=of Y1, yshift=2cm, xshift=1.5cm] (Xmid1) {};
    \node[mynode, left=of Xmid1, xshift=0.75cm] (X21) {Depression};
    \node[mynode, left=of X21,xshift=-0cm] (X11) {Hypertension};
    \node[mynode, right=of Xmid1, xshift=-0.5cm] (X-11) {Sex};
    \node[mynode, right=of X-11,xshift=0cm, minimum width = 2.5cm] (Xn1) {SleepDuration};
    \node[obs, below=of X11, minimum size = 1cm] (M1) {$M_1$};
    \node[obs, below=of X-11, minimum size = 1cm] (M2) {$M_2$};

    \edge {M1,M2} {Y1};
    \edge {X21,X-11,Xn1} {M2};
    \edge{X11} {M1};
    \end{tikzpicture}}
    \caption{Optimal SICI model structure for the Anxiety node in the Cardiovascular BN \cite{Ordovas2023}}
    \label{fig:OptSICIAnxiety}
\end{figure}

The optimal SICI model features the partition of the parents into the two blocks seen by the structure in Figure \ref{fig:OptSICIAnxiety}. The combination function was optimised for this structure through searching the space of partitions of the four potential configurations of $\mathbf{m}$ into two blocks, resulting in a function for which `Anxiety = Yes' if and only if $M_1=1$ and $M_2=0$. This function can only be interpreted via an evaluation of the quantitative parameters themselves; the algorithm may find the optimal parameters such that $M_i$ has a positive or a negative effect on the child, and this affects the optimisation and the interpretation of the partition of the configurations $\mathbf{m}$ that defines $f$. In this case, the presence of hypertension increases the probability that $M_1=1$, in turn allowing anxiety to be present. The larger CPT defining $M_2$ features high probabilities for `$M_2 = 1$' in contrast to the lower probabilities for `$\mathbb{P}(\text{Anxiety = Yes})$' in the original CPT, indicating that the parameters have been defined such that $M_2=0$ supports the presence of anxiety. This would in turn allow us to interpret the output of the algorithm as an AND gate between the two sets of risk factors rather than an OR or XOR gate, for example. The approximation of the original CPT $Y|\mathbf{X}$ can then be constructed through the equation found in Section \ref{SICI}. 

The optimal SICI model CPT, whose parameters are presented in Table \ref{tab:CPTResults}, features 14 free parameters and produces a score of 0.3700 (4dp).

\subsection{Results}

Table \ref{tab:Results} summarises the performance of each structural method for the modelling of the Anxiety node in the Cardiovascular BN \cite{Ordovas2023}. For each method, the optimal, lowest-scoring model found is reported, featuring the score itself (by sum of row-wise total variation distances) and the parameter savings it brings against the full CPT of $Y|\mathbf{X}$. This table omits complexities associated with any choice of structure or deterministic combination function, though these factors are discussed later. When constructing a model through expert judgement, the most burdensome, unintuitive and error-prone aspect for the domain experts is specifying each of the probabilities (i.e. the quantitative parameters). We report the quantitative parameter savings to measure the anticipated reduction in difficulty associated with parameterising each refined structure.

\begin{table}[h!]
\centering
\caption{Comparison of lowest score by sum of row-wise total variation distances and parameter savings achieved by each method}
\vspace{0.2cm}
\scalebox{0.85}{
\begin{tabular}{l|c|c|c|c|c}
\toprule
Method & Pruning & Divorcing & SCMs & ICI & SICI \\
\midrule
Optimal Score (4dp)        & 0.6487 & 0.5072 & 1.2693 & 0.5520 & 0.3700 \\
Number of Parameters  & 12     & 8      & 2      & 9      & 14     \\
Parameter Savings     & 12     & 16     & 22     & 15     & 10     \\
\bottomrule
\end{tabular}
}
\label{tab:Results}
\end{table}

Unsurprisingly, the simple canonical model performs the worst by a significant margin. The benefit of an SCM is that it requires just $s_c-1$ probabilities to model a child that has $s_c$ states, independent of the original parent set $\mathbf{X}$. This comes at the cost of greatly reduced flexibility compared to other structural methods. The number of possible combination functions $f$ grows exponentially in the number of parents $n$, making it even harder to determine which function best represents the system at hand. However, as it only requires a very small number of quantitative parameters, it could be relatively quick to attempt to construct an SCM through discussions with domain experts. If the real-world system features a number of deterministic or almost-deterministic relationships, it could be worth exploring the use of an SCM. In most cases, more sophisticated methods will be needed to account for a more flexible, stochastic representation of the interacting effects of the parents on the child.

Pruning is the second-worst performing method - albeit with a significantly better score than SCMs. Pruning is a very quick way to reduce the parameter space of a local BN structure, and can be performed easily in practice. Domain experts will find it relatively easy to compare the relative influence of each parent. Indeed, this is a vital component in many expert elicitation methodologies \cite[e.g.][]{Cainsmethod,Hassall2019,Phillipson2021,Roed2009}. We can simply elicit these relative influence scores to determine which parents are suitable for pruning. Depending on the size of the original CPT, pruning may still leave a large approximate CPT, with each probability conditional on a number of factors. As a result, it can remain a challenge to parameterise this approximate CPT whether using expert judgement or data. More complex structural refinement approaches not only possibly further reduce the number of quantitative judgements required, but they typically reduce the number of conditioning variables within each conditional probability to be elicited. Pruning can be seen to perform fairly well for the Anxiety node, but it should only be considered a viable option if the domain experts score the relative influence of one or more variables sufficiently low. If a moderate number of parents remain after pruning, this approach will not be sufficient on its own for efficiently approximating the local structure. Care must be taken not to prune any influential parents as it poses a threat of high information loss (which is hard to detect without knowledge of the true CPT), and this may lead to a poor approximation of the local system.

Divorcing outperforms pruning for the modelling of the Anxiety node, having a lower optimal score and a higher parameter saving. This is not surprising as it reduces the number of parameters needing to be determined while explicitly allowing simple interactions between a small subset of parents. This gives divorcing techniques a good level of flexibility, and a good balance between complexity and efficiency. In practice, it should be relatively simple for domain experts to determine whether there are any small subsets of parents that can be suitably modelled through (a series of) logic gates. This can generally be determined through natural language discussions rather than through probabilistic judgements - ensuring the process is accessible and intuitive for the experts. That said, for real-world systems featuring more complex interdependencies, it may become necessary to divorce a greater number of parents. This makes it much harder for the domain experts to find a satisfactory deterministic operator that combines each of the divorced parents into the intermediate node. Furthermore, even for small subsets of divorced parents, assuming the divorced parents interact in such a locally deterministic way may be overly simplistic. Overall, divorcing is a simple yet effective approach that should be considered especially when small subsets of parents with relative simple interdependencies can be identified. The divorcing process relies on structural rather than probabilistic judgements, hence it can be attempted without much cost if it later transpires that divorcing is unsuitable.

The final two methods, ICI and SICI, extend the divorcing methodology across the whole parent set. Both the ICI and SICI models outperform the divorcing methodology on the optimal score found, though not on parameter savings. This is expected as both methodologies are more flexible but more structurally complex. The ICI methodology does require a complex deterministic combination function to be defined over the set of $n$ mechanisms, posing a similar challenge to SCMs. While it improves significantly over SCMs by introducing stochasticity before this deterministic combination takes place rather than after, eliciting or learning a satisfactory combination function for the ICI model can still be difficult. If an AND, OR or XOR gate proves satisfactory, this will likely be found by the expert. However, there are a very large number of possible deterministic combination functions that can be defined through compositions of Boolean operators, and the domain experts may struggle to identify and evaluate more complicated such functions. ICI is expected to perform well in cases where there are very limited interactions between parents, and when there is an approximately rule-based system, embedded with some stochasticity in its inputs, that determines the value of the quantity of interest. When the interactions present are not so simple, the ICI model may be too rigid, and other, more flexible techniques should be explored.

The SICI model \textit{does} explicitly represent interactions between particular subset of parents, giving it greater flexibility than the ICI model. It also encourages the use of fewer intermediate mechanism nodes than the ICI model, reducing the space of combination functions defining $Y|\mathbf{M}$ from which the expert must specify their best option. In defining $\mathbf{M}|\mathbf{X}$, the SICI model features a number of deterministic combination functions, but each is typically defined on a small number of parents. A domain expert may find it simple to define such functions on just two or three parents, though may struggle when this number of parents increases any higher. The SICI model scores better than the ICI model - as expected for a generalisation of the ICI model, though this comes with the cost of reduced parameter savings. As this model is more complex than other structural approaches, other options should be explored first, unless the domain is known to feature complex interaction structures. In particular, it is advisable to establish that the ICI model is too restrictive before adopting the SICI model. Both the ICI and SICI methodologies appear to be relatively good options here for modelling the Anxiety node, with SICI providing the best model of all by row-wise total variation distance against the true CPT.

\section{Discussion} \label{Discussion}

This paper reviews a selection of methods for refining a local structure within a Bayesian network to facilitate efficient model parameterisation. Each method enforces particular, distinct structural restrictions around a node whose CPT is being parameterised, enabling an approximate CPT to be defined through a reduced number of quantitative parameters. These structural methods provide an alternative approach to purely quantitative methods such as those based on interpolation and regression, though both approaches can, and often should, be used in combination. Such a modeller should consult literature on quantitative methods for CPT approximation \cite[e.g.][]{Blomaard2025,Mkrtchyan2016} as well as this paper in order to decide the best approach to take for their modelling problem.

These approximation methods address a significant challenge within Bayesian network modelling, and provide an avenue for the adoption of Bayesian networks in domains that have so far been unable or unwilling to adopt them as a modelling technique. Without a vast quantity of high-quality data available, Bayesian network parameterisation can be very challenging. When using data that is not of sufficient quality and quantity, the resulting parameter estimates may be unstable and unreliable \cite{Rohmer2020}. If formal expert judgement elicitation is used, the process will be lengthy and costly, with experts struggling to provide accurate probabilistic assessments, particularly once fatigue begins to set in \cite{Burgman2015}. In both cases, reducing the parameter space through either a structural or quantitative approach (or both together) is an approach that should be considered. This would make the parameterisation of the network not only more efficient, but possibly even more faithful. The reduced model flexibility resulting from the reduced model parameter space may be countered by a higher level of engagement and reduced fatigue from domain experts providing judgements, or, alternatively, by more stable and reliable estimates for those nodes for which only some data is available. 

Our review not only discusses the foundations of each method, but provides an evaluation through a worked example of how each method performs against a CPT that has been learnt from a large dataset. We have discussed the performance of each method in Section \ref{Comparison}. For our worked example, the SICI model provided the best fit, but also brought the lowest parameter savings. The ICI model provided an adequate fit, but was limited in capturing the interactions among the parent set compared to the SICI model. Divorcing performed well, and is a highly practicable approach to reducing a local structure's parameter space without removing information altogether. Pruning is an effective method for reducing a parameter space, though should be used sparingly only when parents of low influence can be identified. The SCM provided the greatest parameter savings, but provided the worst fit by far. This is a likely outcome when using an SCM due to its rigidity and reliance on a deterministic combination function over the entire parent set. 

In brief, we recommend evaluating whether any parents of low influence can be immediately pruned from the system without notable information loss before evaluating which further approach should be taken, if still necessary. An SCM is highly unlikely to provide a good fit unless the system is naturally highly deterministic, and is thus not generally recommended. Divorcing techniques, including ICI and SICI, should be considered in most cases. Divorcing in general provides a relatively quick and notable parameter saving, while nonetheless allowing many interactions between particular parents to remain unconstrained. ICI and SICI extend general divorcing approaches by considering how the entire range of parents can be partitioned according to their causal mechanisms, and thus may provide a better fit. ICI should be explored first to evaluate the cost of its independence assumption. If this cost is low, the parameter savings are likely worthwhile. If, however, this cost is high and the model is suppressing important interactions between parents, the SICI model may be recommended to express these interactions. This would result in reduced parameter savings, though this may be necessary for obtaining an approximation that is truly faithful to expert beliefs about the real-world system. At this point, we would recommend exploring available quantitative approximation methods that could be incorporated into the model to ease the quantitative elicitation burden while retaining sufficient flexibility. The above recommendations may lead to different approaches being appropriate for different nodes even within the same network. The choice of approach should be made on a node-by-node basis, evaluating the characteristics of structural, quantitative and combined approaches as there is no universally optimal approach. 

These structural methods may be useful not only for efficient model parameterisation, but also for model explainability. Any modification of the local structure around a node can be shown to different stakeholders, including clients who benefit from intuitive visual explanations, auditors who demand rationale at each stage of the model development process, and other domain experts who can interpret any explicitly represented causal mechanisms introduced. The same logic used by domain experts to refine the structure of the network can then be presented to these stakeholders. While the parameters obtained can be interpreted within the original network structure, presenting the modified graphical structure provides a much more intuitive and engaging explanation to stakeholders than the numbers alone. Each modification should be well documented, but does not necessarily have to be displayed as part of the final model; once the final CPTs have been determined, the original, unmodified structure may be displayed to clients and other stakeholders for brevity, but the modified local structures must be documented and stored to help explain each component of the model to stakeholders.

While this review provides practical insight into a variety of structural CPT approximation methods that facilitate efficient BN parameterisation, it does not, and cannot, provide precise criteria as to when each method should be used. Such precise, formulaic criteria may well be impossible to develop without having data to characterise the properties of the local structure being modelled. The most suitable method to use may further depend on the resources available to the modeller such as the timeframe for model development, the amount of data that \textit{is} available for a given node, and the number of experts that are available to provide their judgements. There is no universally optimal solution, and we are continuing to research this problem to provide sound, practical advice to Bayesian network modellers working on real-world problems suffering from data insufficiency. In particular, we are planning to construct a new Bayesian network model for an application in which we are already building a model using a full, formal elicitation protocol \cite{Hanea2017IDEA}. We aim to have a fully elicited, complete model for this application in the coming months. This will act as a benchmark for a new Bayesian network model utilising a variety of approximation techniques found in this review and elsewhere \cite[e.g][]{Blomaard2025,Mkrtchyan2016}. We will then be able to provide further insight - resulting from real-world applications of these approximation methodologies - as to when each approximation method may be suitable. This will also draw in a much larger range of nodes that will each have their own characteristics and thus differing optimal approximation methods. This initial review provides a basis for developing more comprehensive guidance for practical Bayesian network modelling going forward.




\bibliographystyle{elsarticle-num}
\bibliography{NEWBIB}

\newgeometry{margin=2cm}
\appendix

\section{CPT Approximations}
\vspace{-1.5cm}
\begin{table}[H]
\centering
\rotatebox{90}{
    \begin{minipage}{\textheight}
    \centering
    \scriptsize
    \caption{The true CPT parameters against the optimal parameters found for each method, alongside each corresponding optimal score}
    \label{tab:CPTResults}
    \vspace{0.3cm}
\begin{tabular}{@{}l|l|l|l|l|l|l|l|l|l|l|l|l|l|l|l|l@{}}
\toprule
    \multicolumn{5}{c|}{}          & \multicolumn{2}{c|}{Anxiety CPT} & \multicolumn{2}{c|}{Edge Pruning}  & \multicolumn{2}{c|}{Divorcing}     & \multicolumn{2}{c|}{SCM}           & \multicolumn{2}{c|}{ICI}           & \multicolumn{2}{c}{SICI}          \\ \midrule
   & Dep. & Hyp. & Sex    & SD        & No            & Yes          & No                       & Yes    & No                       & Yes    & No                       & Yes    & No                       & Yes    & No                       & Yes    \\
1  & No         & No           & Female & 6-9hours             & 0.9630        & 0.0370       & 0.9730                   & 0.0270 & 0.9434                   & 0.0566 & 0.9393                   & 0.0607 & 0.9642                   & 0.0358 & 0.9606                   & 0.0394 \\
2  & Yes        & No           & Female & 6-9hours             & 0.9830        & 0.0170       & 0.9730                   & 0.0270 & 0.9737                   & 0.0263 & 0.9393                   & 0.0607 & 0.9823                   & 0.0177 & 0.9822                   & 0.0178 \\
3  & No         & Yes          & Female & 6-9hours             & 0.9147        & 0.0853       & 0.9375                   & 0.0625 & 0.9434                   & 0.0566 & 0.9393                   & 0.0607 & 0.9191                   & 0.0809 & 0.9067                   & 0.0933 \\
4  & Yes        & Yes          & Female & 6-9hours             & 0.9603        & 0.0397       & 0.9375                   & 0.0625 & 0.9737                   & 0.0263 & 0.9393                   & 0.0607 & 0.9557                   & 0.0443 & 0.9579                   & 0.0421 \\
5  & No         & No           & Male   & 6-9hours             & 0.8764        & 0.1236       & 0.8794                   & 0.1206 & 0.8750                   & 0.1250 & 0.9393                   & 0.0607 & 0.8679                   & 0.1321 & 0.9301                   & 0.0699 \\
6  & Yes        & No           & Male   & 6-9hours             & 0.8825        & 0.1175       & 0.8794                   & 0.1206 & 0.7955                   & 0.2045 & 0.9393                   & 0.0607 & 0.8994                   & 0.1006 & 0.8878                   & 0.1122 \\
7  & No         & Yes          & Male   & 6-9hours             & 0.8409        & 0.1591       & 0.7955                   & 0.2045 & 0.8750                   & 0.1250 & 0.7500                   & 0.2500 & 0.7631                   & 0.2369 & 0.8342                   & 0.1658 \\
8  & Yes        & Yes          & Male   & 6-9hours             & 0.7500        & 0.2500       & 0.7955                   & 0.2045 & 0.7955                   & 0.2045 & 0.7500                   & 0.2500 & 0.7538                   & 0.2462 & 0.7341                   & 0.2659 \\
9  & No         & No           & Female & \textless{}6hours    & 0.9506        & 0.0494       & 0.9643                   & 0.0357 & 0.9434                   & 0.0566 & 0.9393                   & 0.0607 & 0.9676                   & 0.0324 & 0.9683                   & 0.0317 \\
10 & Yes        & No           & Female & \textless{}6hours    & 0.9781        & 0.0219       & 0.9643                   & 0.0357 & 0.9737                   & 0.0263 & 0.9393                   & 0.0607 & 0.9845                   & 0.0155 & 0.9788                   & 0.0212 \\
11 & No         & Yes          & Female & \textless{}6hours    & 0.9352        & 0.0648       & 0.9544                   & 0.0456 & 0.9434                   & 0.0566 & 0.9393                   & 0.0607 & 0.9301                   & 0.0699 & 0.9248                   & 0.0752 \\
12 & Yes        & Yes          & Female & \textless{}6hours    & 0.9737        & 0.0263       & 0.9544                   & 0.0456 & 0.9737                   & 0.0263 & 0.9393                   & 0.0607 & 0.9619                   & 0.0381 & 0.9498                   & 0.0502 \\
13 & No         & No           & Male   & \textless{}6hours    & 0.9239        & 0.0761       & 0.9133                   & 0.0867 & 0.8750                   & 0.1250 & 0.9393                   & 0.0607 & 0.8741                   & 0.1259 & 0.9426                   & 0.0574 \\
14 & Yes        & No           & Male   & \textless{}6hours    & 0.9026        & 0.0974       & 0.9133                   & 0.0867 & 0.7955                   & 0.2045 & 0.9393                   & 0.0607 & 0.9108                   & 0.0892 & 0.8949                   & 0.1051 \\
15 & No         & Yes          & Male   & \textless{}6hours    & 0.8750        & 0.1250       & 0.8125                   & 0.1875 & 0.8750                   & 0.1250 & 0.9393                   & 0.0607 & 0.7929                   & 0.2071 & 0.8640                   & 0.1360 \\
16 & Yes        & Yes          & Male   & \textless{}6hours    & 0.7500        & 0.2500       & 0.8125                   & 0.1875 & 0.7955                   & 0.2045 & 0.7500                   & 0.2500 & 0.7878                   & 0.2122 & 0.7509                   & 0.2491 \\
17 & No         & No           & Female & \textgreater{}9hours & 0.9434        & 0.0566       & 0.9576                   & 0.0424 & 0.9434                   & 0.0566 & 0.9393                   & 0.0607 & 0.9390                   & 0.0610 & 0.8738                   & 0.1262 \\
18 & Yes        & No           & Female & \textgreater{}9hours & 0.9719        & 0.0281       & 0.9576                   & 0.0424 & 0.9737                   & 0.0263 & 0.9393                   & 0.0607 & 0.9665                   & 0.0335 & 0.9583                   & 0.0417 \\
19 & No         & Yes          & Female & \textgreater{}9hours & 0.7000        & 0.3000       & 0.8054                   & 0.1946 & 0.7000                   & 0.3000 & 0.7500                   & 0.2500 & 0.8375                   & 0.1625 & 0.7008                   & 0.2992 \\
20 & Yes        & Yes          & Female & \textgreater{}9hours & 0.9107        & 0.0893       & 0.8054                   & 0.1946 & 0.9107                   & 0.0893 & 0.9393                   & 0.0607 & 0.9098                   & 0.0902 & 0.9013                   & 0.0987 \\
21 & No         & No           & Male   & \textgreater{}9hours & 0.8299        & 0.1701       & 0.8127                   & 0.1873 & 0.8750                   & 0.1250 & 0.7500                   & 0.2500 & 0.8216                   & 0.1784 & 0.7992                   & 0.2008 \\
22 & Yes        & No           & Male   & \textgreater{}9hours & 0.7955        & 0.2045       & 0.8127                   & 0.1873 & 0.7955                   & 0.2045 & 0.7500                   & 0.2500 & 0.8140                   & 0.1860 & 0.8016                   & 0.1984 \\
23 & No         & Yes          & Male   & \textgreater{}9hours & 0.5000        & 0.5000       & 0.5000                   & 0.5000 & 0.5000                   & 0.5000 & 0.7500                   & 0.2500 & 0.5407                   & 0.4593 & 0.5241                   & 0.4759 \\
24 & Yes        & Yes          & Male   & \textgreater{}9hours & 0.5000        & 0.5000       & 0.5000                   & 0.5000 & 0.5000                   & 0.5000 & 0.7500                   & 0.2500 & 0.5000                   & 0.5000 & 0.5296                   & 0.4704 \\ \midrule
  \multicolumn{7}{c|}{} & \multicolumn{2}{c|}{STVD = 0.6487} & \multicolumn{2}{c|}{STVD = 0.5072} & \multicolumn{2}{c|}{STVD = 1.2693} & \multicolumn{2}{c|}{STVD = 0.5520} & \multicolumn{2}{c}{STVD = 0.3700} \\ \bottomrule
\end{tabular}
\end{minipage}
}
\end{table}
\restoregeometry
\end{document}